\documentstyle[12pt,psfig]{article}

\catcode`\@=11
\def\marginnote#1{}
\newcount\hour
\newcount\minute
\newtoks\amorpm
\hour=\time\divide\hour by60
\minute=\time{\multiply\hour by60 \global\advance\minute by-
\hour}
\edef\standardtime{{\ifnum\hour<12 \global\amorpm={am}%
    \else\global\amorpm={pm}\advance\hour by-12 \fi
    \ifnum\hour=0 \hour=12 \fi
    \number\hour:\ifnum\minute<100\fi\number\minute\the\amorpm}}
\edef\militarytime{\number\hour:\ifnum\minute<100\fi\number\minute}
\def\draftlabel#1{{\@bsphack\if@filesw {\let\thepage\relax
  \xdef\@gtempa{\write\@auxout{\string
    \newlabel{#1}{{\@currentlabel}{\thepage}}}}}\@gtempa
    \if@nobreak \ifvmode\nobreak\fi\fi\fi\@esphack}
     \gdef\@eqnlabel{#1}}
\def\@eqnlabel{}
\def\@vacuum{}
\def\draftmarginnote#1{\marginpar{\raggedright\scriptsize\tt#1}}
\def\draft{\oddsidemargin -.5truein
        \def\@oddfoot{\sl preliminary draft \hfil
        \rm\thepage\hfil\sl\today\quad\militarytime}
        \let\@evenfoot\@oddfoot \overfullrule 3pt
        \let\label=\draftlabel
        \let\marginnote=\draftmarginnote

\def\@eqnnum{(\theequation)\rlap{\kern\marginparsep\tt\@eqnlabel}%
\global\let\@eqnlabel\@vacuum}  }
\def\preprint{\twocolumn\sloppy\flushbottom\parindent 1em
        \leftmargini 2em\leftmarginv .5em\leftmarginvi .5em
        \oddsidemargin -.5in    \evensidemargin -.5in
        \columnsep 15mm \footheight 0pt
        \textwidth 250mmin      \topmargin  -.4in
        \headheight 12pt \topskip .4in
        \textheight 175mm
        \footskip 0pt

\def\@oddhead{\thepage\hfil\addtocounter{page}{1}\thepage}
        \let\@evenhead\@oddhead \def\@oddfoot{} \def\@evenfoot{}
}
\def\titlepage{\@restonecolfalse\if@twocolumn\@restonecoltrue\onecolumn
     \else \newpage \fi \thispagestyle{empty}\c@page\z@
        \def\thefootnote{\fnsymbol{footnote}} }
\def\endtitlepage{\if@restonecol\twocolumn \else  \fi
        \def\thefootnote{\arabic{footnote}}
        \setcounter{footnote}{0}}  
\catcode`@=12
\relax
\def\be{\begin{equation}}
\def\ee{\end{equation}}
\def\bea{\begin{eqnarray}}
\def\eea{\end{eqnarray}}
\def\simlt{\stackrel{<}{{}_\sim}}
\def\simgt{\stackrel{>}{{}_\sim}}

\def\NPB#1#2#3{{\it Nucl.~Phys.} {\bf{B#1}} (19#2) #3}
\def\PLB#1#2#3{{\it Phys.~Lett.} {\bf{B#1}} (19#2) #3}
\def\PRD#1#2#3{{\it Phys.~Rev.} {\bf{D#1}} (19#2) #3}
\def\PRL#1#2#3{{\it Phys.~Rev.~Lett.} {\bf{#1}} (19#2) #3}
\def\ZPC#1#2#3{{\it Z.~Phys.} {\bf C#1} (19#2) #3}

\def\PR#1#2#3{{\it Phys.~Rep.} {\bf#1} (19#2) #3}
\def\RMP#1#2#3{{\it Rev.~Mod.~Phys.} {\bf#1} (19#2) #3}
\def\HPA#1#2#3{{\it Helv.~Phys.~Acta} {\bf#1} (19#2) #3}

\def\mst1{m_{\widetilde{t}_1}}
\def\mst2{m_{\widetilde{t}_2}}
\def\mst12{m_{\widetilde{t}_{1,2}}}

\def\msb1{m_{\widetilde{b}_1}}
\def\msb2{m_{\widetilde{b}_2}}
\def\msb12{m_{\widetilde{b}_{1,2}}}

\def\mtilde2{\widetilde{m}^{2}}

\relax

\begin{document}
\setlength{\baselineskip}{3.0ex}
\begin{titlepage}
\phantom{bla}
\begin{flushright}
CERN-TH/95-197\\
IEM-FT-110/95 \\
hep--ph/9507317 \\
\end{flushright}
\vskip 0.3in
\begin{center}{\bf
WOULD A LIGHT HIGGS DETECTION IMPLY NEW PHYSICS?
\footnote{Based on talk given at {\it Electroweak interactions and unified
theories}, Les Arcs, France, 11-18 March 1995.}} \\  \vspace{2cm}
{\bf M. Quir\'os} \footnote{Work supported in part by
the European Union (contract CHRX-CT92-0004) and
CICYT of Spain
(contract AEN94-0928).}
\\
CERN, TH Division, CH--1211 Geneva 23, Switzerland\\
\end{center}
\vspace{2.5cm}

\vskip.5cm
\begin{center}
{\bf Abstract}
\end{center}
\vbox{ \baselineskip 14pt
Depending on the Higgs-boson and  top-quark masses,
$M_H$ and $M_t$, the effective potential of the Standard
Model can develop a non-standard minimum for values of
the field much larger than the weak scale. In those cases the
standard minimum becomes metastable and the
possibility of decay to the non-standard one arises.  Comparison of
the decay rate to the non-standard minimum at finite (and
zero) temperature with the corresponding expansion rate
of the Universe
allows to identify the region, in the ($M_H$,
$M_t$) plane, where the Higgs field is sitting at the standard
electroweak minimum. Since that region depends on the cutoff
scale $\Lambda$, up to which we believe the Standard Model,
the discovery of the Higgs boson, mainly at LEP-200, might
put an upper bound (below the Planck scale)
on the scale of new physics $\Lambda$.
}
\vspace{5cm}

\begin{flushleft}
CERN-TH/95-197\\
July 1995 \\
\end{flushleft}

\end{titlepage}
\setcounter{footnote}{0}
\setcounter{page}{0}
\newpage
\section{Introduction}
For particular values of the Higgs boson and top quark masses, $M_H$ and $M_t$,
the effective potential of the Standard Model (SM) develops a deep non-standard
minimum for values of the field $\phi \gg G_F^{-1/2}$ \cite{L}.
In that case the
standard electroweak (EW) minimum becomes metastable and might decay into
the non-standard one. This means that the SM might not accomodate
certain regions
of the plane ($M_H$,$M_t$), a fact
which can be intrinsically interesting as evidence for
new physics. Of course, the mere existence of the non-standard minimum,
and also
the decay rate
of the standard one into it, depends on the scale $\Lambda$ up to which
we believe the SM results. In fact, one can identify $\Lambda$
with the scale of new
physics.

In this talk I will review the present situation on the above issue
and its relevance for
evidence of new physics if a light Higgs is detected experimentally,
most likely at
LEP-200.

\section{When the EW minimum becomes metastable?}
The preliminary question one should ask is: When the standard EW minimum
becomes
me\-ta\-sta\-ble, due to the appearance of a deep non-standard
minimum? This question was
addressed in past years \cite{L} taking into account leading
log and part of next-to-leading
log  corrections. More recently, calculations have
incorporated all next-to-leading log
corrections \cite{AI,CEQ}. In particular in ref.~\cite{CEQ}
next-to-leading  log corrections
are resummed to all-loop by the renormalization group equations (RGE),
and pole masses for the top-quark and
the Higgs-boson are considered.
{}From the requirement of a stable (not metastable) standard EW
minimum we obtain a lower bound on
the Higgs mass, as a function of the top mass, labelled by
the values of the SM cutoff
(stability bounds). Our
result \cite{CEQ} is lower than previous estimates by ${\cal O}$(10) GeV.

The one-loop effective potential of the SM improved by two-loop
RGE has been shown to
be highly scale independent \cite{CEQR} and, therefore, very
reliable for the present study.
It has stationary points at
\be
\label{minimo}
\phi^2 = \frac{2m^2}{\widetilde{\lambda}}; \ \
\widetilde{\lambda} = \lambda-\frac{3}{8\pi^2}h_t^4
\left(\log\frac{h_t^2}{2}-1\right)
\ee
where $m^2$ and $\lambda$ are the tree-level mass and
quartic coupling parameters of the SM, and $h_t$ is the
top-quark Yukawa coupling. All parameters in (\ref{minimo})
are running with the renormalization
scale, that has been identified with the field $\phi$,
and we are keeping, to simplify the presentation,
only the top-quark Yukawa coupling in the one-loop correction.

The second derivative of the effective potential at (\ref{minimo}) is
\be
\label{curvatura}
V''(\phi)=2m^2+\frac{1}{2}\beta_{\lambda}\phi^2
\ee
A quick glance at (\ref{minimo}) shows that eq.~(\ref{minimo})
can be satisfied for values of the
field $\phi\gg v=246.22$ GeV, provided that,
for those values of the field, $\widetilde{\lambda}\sim 0$.
In this case, since $m^2\sim 10^2$ GeV for all values of the scale,
the first term in
(\ref{curvatura}) is negligible and the second term will control
the nature of the stationary point.
In particular,
\bea
\beta_{\lambda} < 0  & \Longrightarrow & V''< 0  \  \ {\rm (MAXIMUM)}
\nonumber \\
\beta_{\lambda} > 0  & \Longrightarrow & V''> 0  \  \ {\rm (MINIMUM)}
\nonumber
\eea

In Fig.~1 we show (thick solid line) the shape of the effective
potential for $M_t=175$ GeV
and $M_H=121.7$ GeV. We see the appearance of the non-standard maximum,
$\phi_M$, while the global
non-standard minimum has been cutoff at $M_P$. We can see from
Fig.~1 the steep descent from
the non-standard maximum. Hence, even if the non-standard minimum
is beyond the SM cutoff, the
standard minimum becomes metastable and can be destabilized.
So for fixed values of $M_H$ and
$M_t$ the condition for the standard minimum not to become metastable is
\be
\label{condstab}
\phi_M \simgt \Lambda
\ee
Condition (\ref{condstab}) makes the stability condition $\Lambda$-dependent.
In fact we have plotted
in Fig.~2 the stability condition on $M_H$ versus $M_t$ for two
different values of $\Lambda$,
$10^{19}$ GeV (left panel) and 10 TeV (right panel). In both figures
the stability region corresponds to
the region above the dashed curves.

\begin{figure}[hbt]
\centerline{
\psfig{figure=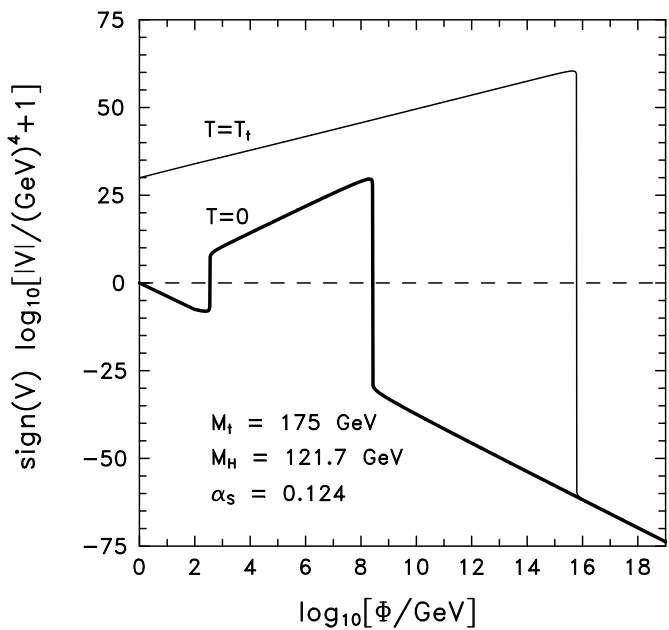,height=8.5cm,bbllx=6.cm,bblly=3.cm,bburx=15.5cm,bbury=16cm}}
\caption{Plot of the effective potential for $M_t=175$ GeV, $M_H=121.7$
GeV at $T=0$ (thick solid line) and $T=T_t=2.5\times 10^{15}$ GeV
(thin solid line).}
\end{figure}

\section{When the EW minimum decays?}

In the last section we have seen that in the region of Fig.~2
below the dashed lines the standard EW minimum is
metastable. However we should not draw physical consequences from
this fact since we still do not
know at which minimum does the Higgs field sit. Thus, the
real physical constraint we have
to impose is avoiding
the Higgs field sitting at its non-standard minimum. In fact the
Higgs field can be sitting at its
non-standard minimum at zero temperature because:
\begin{enumerate}
\item
The Higgs field was driven from the origin to the non-standard
minimum at finite temperature
by thermal fluctuations in a non-standard EW phase transition at
high temperature.
This minimum evolves naturally to the non-standard minimum
at zero temperature. In this case
the standard EW phase transition, at $T\sim 10^2$ GeV, will not take place.
\item
The Higgs field was driven from the origin to the standard minimum at
$T\sim 10^2$ GeV, but decays,
at zero temperature, to the non-standard minimum by a quantum fluctuation.
\end{enumerate}

In Fig.~1 we have depicted the effective potential at
$T=2.5\times 10^{15}$ GeV (thin solid line) which
is the corresponding
transition temperature. Our finite temperature potential \cite{EQ}
incorporates plasma effects
\cite{Q} by one-loop resummation of Debye masses \cite{DJW}.
The tunnelling probability per unit time per
unit volume  was computed long ago for thermal \cite{Linde} and
quantum \cite{Coleman} fluctuations.
At finite temperature it is given by $\Gamma/\nu\sim T^4 \exp(-S_3/T)$,
where $S_3$ is the euclidean action evaluated
at the bounce solution $\phi_B(0)$. The semiclassical picture is
that unstable bubbles are nucleated behind the
barrier at $\phi_B(0)$ with a probability given by $\Gamma/\nu$.
Whether or not they fill the Universe depends on
the relation between the probability rate and the expansion rate of the
Universe. By normalizing the former
with respect to the latter we obtain a normalized probability $P$,
and the condition for decay corresponds
to $P\sim 1$. Of course our results are trustable,  and the decay
actually happens, only if
$\phi_B(0)<\Lambda$, so that the similar condition to (\ref{condstab}) is
\be
\label{condmeta}
\Lambda< \phi_B(0)
\ee
\begin{figure}[htb]
\centerline{
\psfig{figure=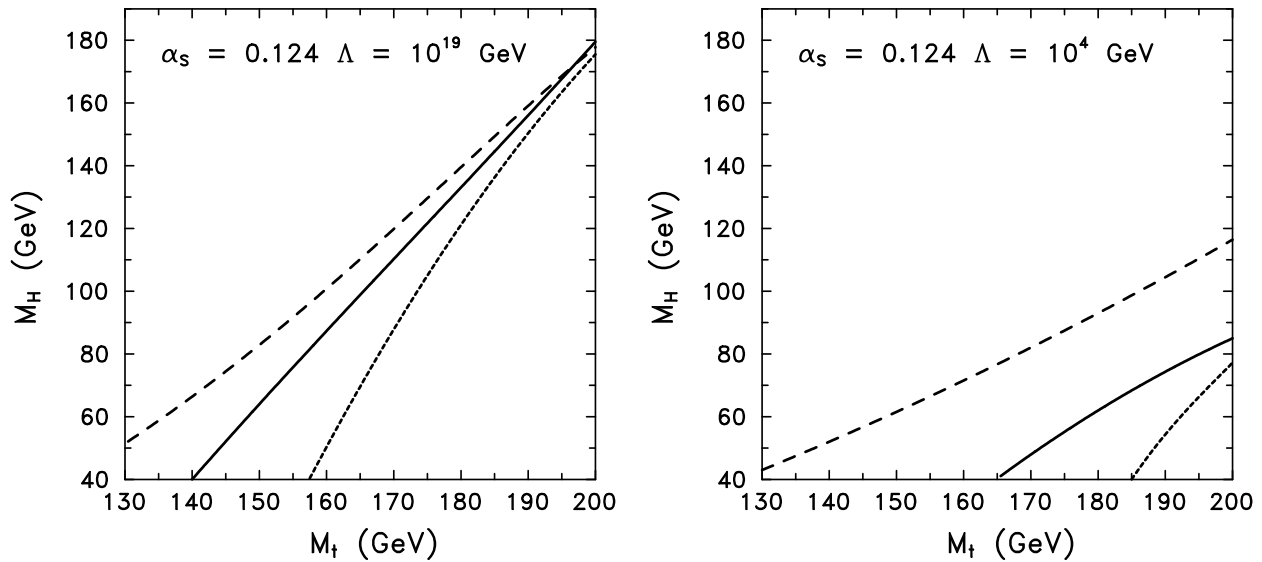,height=9.5cm,bbllx=9.5cm,bblly=1.cm,bburx=19.cm,bbury=14cm}}
\caption{Lower bounds on $M_H$ as a function of $M_t$, for
$\Lambda=10^{19}$ GeV (left panel) and 10 TeV (right panel). The dashed curves
correspond to the stability bounds of Section 2 and  the solid (dotted)
ones to the metastability
bounds of Section 3 at finite (zero) temperature.}
\end{figure}

The condition of no-decay (metastability condition) has been plotted in
Fig.~2  (solid lines)
for the two values of $\Lambda$, $10^{19}$ GeV (left panel) and 10 TeV
(right panel). In both cases the region
between the dashed and the solid line corresponds to a situation where
the non-standard minimum exists
but there is no decay to it at finite temperature. In the region below
the solid lines the Higgs field is sitting
already at the non-standard minimum at $T\sim 10^2$ GeV, and the  standard EW
phase transition does not happen.

We also have evaluated the tunnelling probability at zero temperature from
the standard EW minimum to the
non-standard one. The result of the calculation should translate, as
in the previous case, in lower bounds
on the Higgs mass for differentes values of $\Lambda$. The
corresponding bounds are shown in Fig.~2 in
dotted lines. Since the dotted line lies always below the solid one,
the possibility of quantum tunnelling at
zero temperature does not impose any extra constraint.

As a consequence of all improvements in the calculation, our bounds are
lower than previous estimates
\cite{AV}. To fix ideas, for $M_t=175$ GeV, the bound reduces by $\sim 10 $
GeV for $\Lambda=10^4$ GeV,
and $\sim 30$ GeV for $\Lambda=10^{19}$ GeV.

\section{Does a light Higgs imply new physics?}

{}From the previous discussion it should be clear by now that the Higgs
and top mass measurements
could serve to discriminate between the SM and its extensions, and
to provide information about the
scale of new physics $\Lambda$. In Fig.~3 (left panel) we give the SM
lower bounds on
$M_H$ for $\Lambda\simgt 10^{15}$ (thick lines) and the upper bound
on the mass of the
lightest Higgs boson in the minimal supersymmetric standard model (MSSM)
(thin lines)
for $\Lambda_{\rm SUSY}\sim 1$ TeV. Taking, for instance,  $M_t=180$ GeV,
which coincides with the
central value recently reported by CDF+D0 \cite{top}, and $M_H\simgt 130$
GeV, the SM is
allowed and the MSSM is excluded. On the other hand, if $M_H\simlt 130$ GeV,
then the MSSM is
allowed while the SM is excluded. Likewise there are regions where the
SM is excluded, others
where the MSSM is excluded and others where both are permitted or
both are excluded.
\begin{figure}[htb]
\centerline{
\psfig{figure=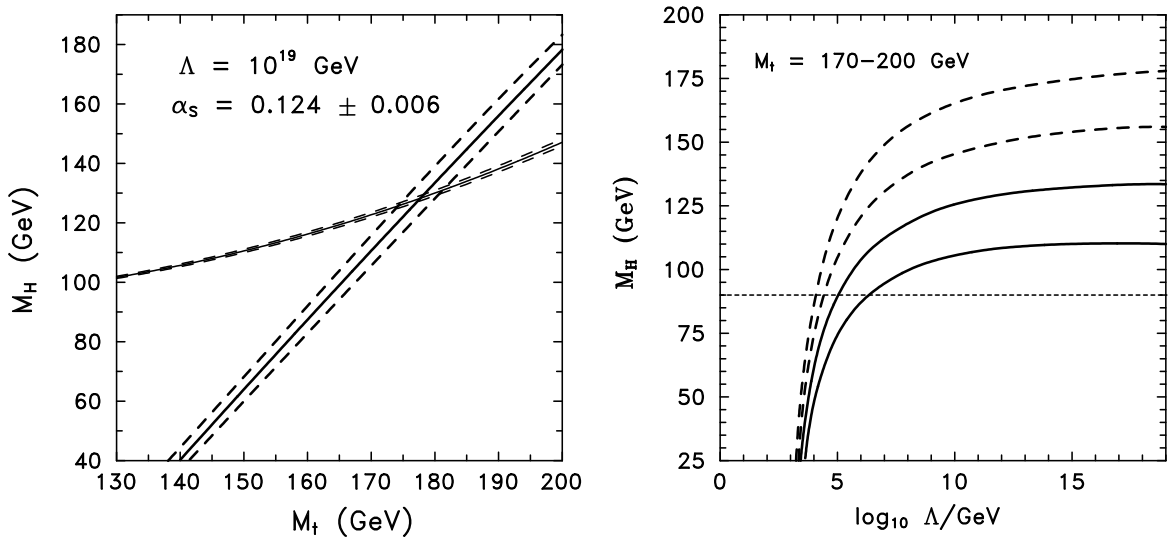,height=9.cm,bbllx=9.5cm,bblly=1.cm,bburx=19.cm,bbury=14cm}}
\caption{Left panel: SM lower bounds on $M_H$ (thick lines) as a function of
$M_t$,
for $\Lambda=10^{19}$ GeV, from
metastability requirements, and upper bounds on the lightest Higgs boson
mass in the
MSSM (thin lines) for $\Lambda_{\rm SUSY}=1$ TeV.
Right panel: SM lower bounds on $M_H$ from
metastability requirements as a function of $\Lambda$ for different
values of $M_t$.}
\end{figure}

Finally from the bounds $M_H(\Lambda)$ (see Fig.~3, right panel) one
can easily deduce that
a measurement of $M_H$ might provide an {\bf upper bound}  (below the
Planck scale) on the
scale of new physics provided that
\be
\label{final}
M_t>\frac{M_H}{2.25\;  {\rm GeV}}+123\; {\rm GeV}
\ee
Thus, the present
experimental bound from LEP, $M_H>64$ GeV, would imply, from
(\ref{final}), $M_t>152$ GeV, which is fulfilled by experimental
detection of the top
\cite{top}. Even non-observation of the Higgs at LEP-200
(i.e. $M_H\simgt 95$ GeV), would
leave an open window ($M_t\simgt 163$ GeV) to the possibility that a
future Higgs detection
at LHC could lead to an upper bound on $\Lambda$. Moreover, Higgs detection at
LEP-200 would put an upper bound on the scale of new physics. Taking,
for instance,  $M_H\simlt 95$
GeV and  170 GeV $< M_t< $ 180 GeV, then $\Lambda\simlt 10^7$ GeV, while for
180 GeV $< M_t <$ 190 GeV, then $\Lambda\simlt 10^4$ GeV, as can be deduced
from
the right panel of Fig.~3.

\end{document}